\documentstyle[12pt,epsf]{article} 
\normalbaselines 
\begin{document} 
\bibliographystyle{unsrt} 
\newcommand{\be}{\begin{equation}}
\newcommand{\ee}{\end{equation}}

\vbox {\vspace{6mm}}  
 
\begin{center}
 
{\large \bf INVARIANTS OF ELEMENTARY OBSERVATION} \\[7mm] 
Johann Summhammer\\ 
{\it Atominstitut\\ Stadionallee 2, A-1020 Vienna, Austria}\\ E-mail: 
summhammer@ati.ac.at\\[5mm]

\end{center} 
 
\vspace{2mm} 
 
\begin{abstract} 
Physics searches for the invariants in the multitude of observations. In this paper we look for 
the invariants of probabilistic observation, whereby we avoid to postulate any physical 
structures as given. We let structure be a consequence of the rarely mentioned assumption of 
science that new information obtained by additional observation should lead to more accurate 
knowledge of the invariants. This leads to unique random variables for expressing 
probabilistic information obtained as clicks in the two detectors of a yes-no experiment: 
Complex probabiliy amplitudes, as familiar from quantum theory. Similarly, the quantum 
mechanical superposition principle is singled out by the demand that also predictions must 
become more accurate with increasing number of observations which serve as input.

By considering that the external conditions of a probabilistic experiment can themselves be 
monitored at the most detailed level, we are lead to probabilistic experiments where two 
clicks happen in coincidence at different detectors. We find that these, and more generally, 
any higher-order coincidence experiments, must be described just as a one-click experiment 
with the same number of possible outcomes.

An observable probability is found to be controllable by two independent experimental 
conditions, which are naturally parametrized by polar coordinates. The relation of a probability 
to the determining conditions is that of an orientation in a 3-dimensional cartesian space.

We conclude that the probabilistic paradigm of current physics inherently defines a method of 
forming concepts and making predictions, which uses the available information optimally, and 
which may be irrefutable within the probabilistic paradigm. Because, whenever a prediction 
turns out wrong, it postulates an as yet unmonitored condition as the culprit, {\it and 
shows how to incorporate it formally}. The Hilbert space formalism of quantum theory 
appears to be isomorphic to this method.
\end{abstract}

\section{Introduction} 
 
In an article on "Physical Reality" Max Born once said that in his opinion the 
key to a reasonable concept of reality, not only in physics, but for all aspects 
of the world, can be found in the idea of the invariant \cite{Born}. In this paper I 
will not discuss "reality", because it is an elusive notion, at best. Rather I will apply Born's 
idea to the basic phenomenology of physics: The observation of "clicks". This phenomenology 
appears to be incapable of further reduction. Therefore, the occurrence of clicks shall be the 
unquestioned starting point from which I will try to extract invariants contained in observed 
data to see what kind of structure emerges. No fields, no particles, no space, let alone any 
specific theory, are to be assumed as given. Indeed, the very idea that the world should be 
made up of "something" with muscular connotation --- whence the term {\it physics}--- will 
be put aside. The structures which arise will solely be due to the interpretative tool for 
extracting meaning from the data of observations.
 
First I want to clarify why I consider the basic phenomenology of physics, which is the 
phenomenology of quantum physics, as irreducible.  In quantum physics an observation yields 
one outcome of a countable number of possible outcomes. In the simplest case only two 
outcomes are possible: A detector may fire or it may not. This was not the case in 
classical physics. There, a pointer could take a continuum of positions. When 
reading the position, the experimenter would in principle have assigned 
values to an infinite number of bits, to put it in modern language. Naturally, he or she could 
not do this 
in finite time, and so had to accept a cut-off error in reading the position of the 
pointer, thereby assigning values to only a {\it finite} number of bits. In quantum 
experiments this comes about automatically. A quantum observation contains no 
cut-off error of reading the instrument. For an unexplained, and perhaps unexplainable reason, 
only a finite number of bits can be observed in finite time \cite{Kant}. 
Now we could ask ourselves whether a still more elementary mode of observation 
is possible. Imagine a hitherto unknown strata of the world. How could it manifest  
itself? Although we cannot exclude that it might affect an 
observer in some unintelligible way, it is clear that, whatever the 
observer {\it records}, and thus captures for further communication, will be 
expressible without loss by a finite number of bits. This means we would be back at the 
phenomenology of quantum physics. 
 
So, with Max Born's view in mind, the task at hand is to look for all kinds of 
invariants that can be extracted from discrete and finite data. In trying to do this, we are 
faced with the need for interpretation at the very beginning. Should we assume that observed 
data are deterministically related to the data of other observations, or should we assume a 
probabilistic link, or something different altogether? I want to discard a deterministic link. 
The reason is that the amount of records available to the observer to form a conception of the 
world is always finite, so that many different sets of laws can be invented to account for 
them. Pinning down any one of these sets as {\it the} laws of nature is then purely 
speculative. On the other hand we have the probabilistic view, which is successfully used to 
interpret quantum observations. It seems that in this view we assign a minimum of 
information content to observed data. To see this, imagine the $N$ trials of a probabilistic 
yes-no experiment, like tossing a coin, in which the outcome "yes" occurs $L$ times. If we 
want to tell somebody else the result it is sufficient to state the values of $N$ and of $L$. 
With the deterministic view, in which the precise sequence of outcomes is important, we 
would in general have to communicate many more details to enable the receiver to 
reconstruct this sequence.

Nevertheless, taking the number of trials and the number of "yes" outcomes as completely 
representing the observed data does not yet constitute the probabilistic view. We could still 
extract $N+1$ different messages from this experiment, because $L$ can be between 0 and 
$N$. The number of different messages would grow linearly with the number of trials. But 
we know that in probabilistic yes-no experiments the number of distinguishable messages 
only grows as the {\it square root} of the number of trials, due to the binomial distribution. 
Of the three views for extracting meaning from $N$ bits the probabilistic view thus expects 
to obtain the least information. This fits well with our premise to assume no physical 
structure as given. Also, the probabilistic view invites the question whether it leads to 
apparant but {\it testable} laws --- even if there are no laws contained in the data --- by 
mirroring to the observer as unexpected structure what he has put in as rules for categorizing 
data. If true, this would indeed be an appealing basis for science.  The possibility of  
lawlessness, and laws being only symmetries, has recently been taken up by several authors 
\cite{Age.Bohr}, \cite{Calude}, \cite{Anandan}, although it may be traced back to 
Kant, and possibly further.

Hence, we shall adopt the probabilistic view. Our aim is to look for invariants in discrete data, 
where we assume that only the probability that we have observed this outcome rather than 
any other one is to render meaning.

\section{The observer's situation}

Before we begin our actual analysis it will be useful to clarify what we mean by observation. 
For us humans observation is the act of taking note of a sensory impression or of a state of 
mind. This ties observation to consciousness, which in turn seems to be separate for different 
individuals. An individual may communicate its observations to others \cite{Shimony}. 
Parallel to the stream of observational data the human observer is aware of the passage of 
time. Without having to disect the concept of time here, I think it is a precondition of science, 
impossible to extract from the discoveries of science. The passage of time can, of course, be 
linked to properties of observable quantities, e.g. the increase of entropy. But ultimately this is 
only a rewording of the {\it a priori} statement "time passes". For our purposes it is 
sufficient to assume that the observer automatically assigns a unique "identity tag" in the form 
of a time stamp to each observed fact. 

The empirical sciences take their input mainly from sensory impressions. These impressions, 
but more generally quantities derived from them, constitute the data which are subjected to 
rational analysis in order to obtain invariants with respect to all kinds of operations. Such 
unveiled invariants are the essential structure contained in the data. Therefore, the world to 
the observer is the set of invariants in {\it his or her data}. 

The role of consciousness in observation is the ability to take note of. This is how data come 
into existence. For an idealized model of the observer we can simply postulate that data come 
into existence at sensors. This liberates our further considerations from the philosophical 
burden of the term consciousness. Hence we can think of the observer as a machine with an 
internal clock, a huge memory to never have to forget data, and equally huge data analysing 
and reasoning abilities \cite{Jaynes}. The machine receives data in the form of a 
persistent stream of bits at a number of sensors. At any unit time interval each of the sensors 
can only show "0" or "1". We could call this machine the {\it common observer} (Fig.1a). A 
further idealization is the assumption that the observer has only one sensor, shown as the 
{\it ultimate observer} in Fig.1b. This is possible, because ultimately the data of each of 
several sensors will have to be taken up by the central calculating and reasoning agent, 
otherwise they cannot be known to that agent and we cannot speak of just one observer.
Neither the common nor the ultimate observer have any possibility to act, because making a 
decision and acting is actually again just a particular stream of data (for us humans it appears 
as data generated internally in our brain and body). The idealized observer has initially no idea 
of "body". This is something it would have to find as a particular set of invariants in the data. 
And it has no concept of existing "in" a space. It should discover space as a convenient form 
of representation of data. All experience of this observer, "external" as well as "internal", 
"mental" as well as "sensory", is just a stream of data. 

We should note that the idea of seeking structure in the mere fact of the persistent increase 
of information in the form of known data is not new. But it seems to have been motivated by 
cosmological considerations, e.g., von Weizs\"acker's {\it ur}-theory \cite{Weizs.}, 
\cite{Drieschner}, \cite{Lyre}, \cite{Lyre2}, or the program which ensued from 
Eddington's {\it Fundamental Theory} \cite{Kilmister}. 

In this paper we will not take direct recourse to either the common or the ultimate observer. 
Nevertheless, I think it is important to illustrate the level of abstraction to which we are lead 
once we accept the {\it empirical finding} of quantum physics that any observation yields 
a discrete and finite outcome.
\begin{figure}[ht]
\begin{center}
\epsffile{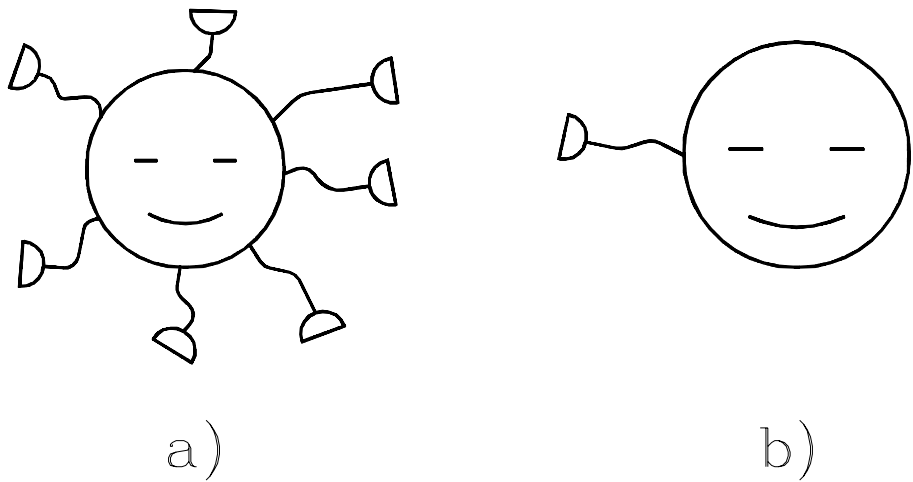}
\caption[Fig.1:]{\small Machine idealizations of the observer. {\bf a)} The {\it 
common observer} has several sensors. {\bf b)} The {\it ultimate observer} has only 
one sensor.}
\end{center}
\end{figure}
What we {\it will} do in the following sections is to investigate the more practical 
question of how we gain information in elementary experiments and what is the optimal way 
to represent this knowledge by means of numbers to bring out the invariants. (Since we do 
not presuppose that nature is in any way "mathematical", the representation of knowledge by 
numbers rather than just arbitrary names will at one point have to be justified. This is not the 
purpose of the paper. We just use numbers as a convenient resource of names. But clearly, 
the fact that a number has two neighbours forces a structure into the representation of 
knowledge, which one should be aware of.) To this end we will characterize experiments 
according to how many separate clicks constitute a single trial. A click is conventionally 
associated with the registration of a particle at a detector which can only indicate click or 
no-click in one unit interval of its time resolution. The simplest case is an experiment where a 
click can occur in one of two detectors (or occurrence or non-occurrence of a click at just 
one detector). This can be extended to experiments with an arbitray number of detectors, 
where the single click of a trial can occur in any of them. The next level of experiments 
would be those, in which the registration of {\it two} clicks constitutes a single trial. Then 
we can go to experiments where three clicks happen per trial, etc. But we will confine 
ourselves to one-click and two-click experiments, because these are sufficient to reveal the 
essential invariants we are aiming for.

Strictly speaking, the distinction of levels of probabilistic experiments according to how many 
clicks constitute a single trial is already a concession to familiar physics. From a logical point 
of view the only relevant criterion is how many different {\it outcomes} are possible in a 
single trial. And if these are more than two, we can always label a particular outcome as 
"yes" and collect the others into "no", so that we have again just yes-no experiments. The 
invariants extractable from the yes-no experiment are therefore basic to {\it all} 
experimentation. We will look at this experiment now.

\section{The observation of a probability}

The aim of observations is to learn something about the situation we are investigating. A more 
correct way of putting this is that we have a number of hypotheses on what the situation 
might be and by means of observations we try to exclude as many of them as we can, so that 
only few remain as possibly true at the chosen level of confidence. An everyday example 
would be the test of the state of charge of a battery. Initially we cannot say anything. 
Therefore the range of hypotheses has to encompass all possibilities from "no charge" to "full 
charge". By measurement of the current through a specified resistance we obtain information. 
On the basis of these data we exclude all hypotheses except those which conjectured a state 
of charge compatible with the uncertainty interval of the result of the measurement.

When we determine the parameters of a physical situation through the measurement of a 
probability, we proceed in the same manner.
For the sake of concreteness, we imagine an experiment where a particle can impinge on one 
of two detectors (Fig.2). The overall physical situation determines the probability that it 
impinges on detector 1, and thereby also the complementary probability that it impinges on 
detector 2. 
\begin{figure}[ht]
\begin{center}
\epsffile{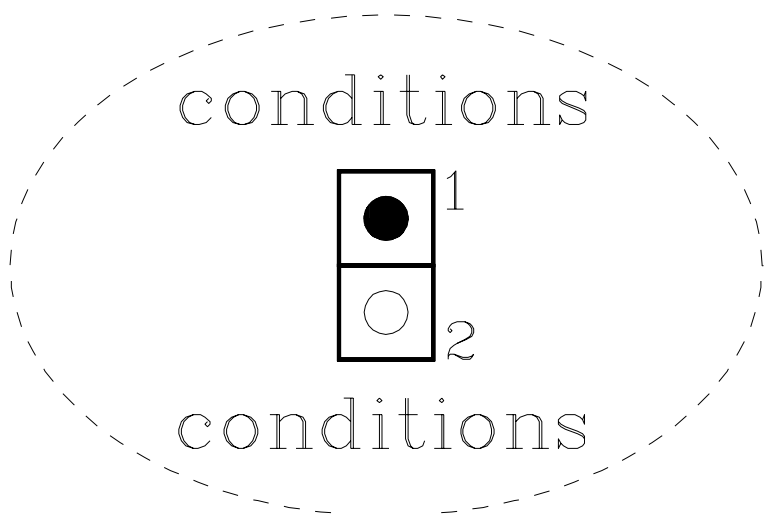}
\caption[Fig.2:]{\small The yes-no experiment. The physical conditions determine the 
probability for a click to happen at detector 1 or at detector 2. The click is indicated by the 
full circle.}
\end{center}
\end{figure}
We assume we do not know which physical conditions have been set.
By measuring the probability $p$ for a click in detector 1 (or, conversely, of $1-p$ for a click 
in detector 2) we wish to eliminate hypotheses on the physical situation. We will denote 
hypotheses by the real variable $\chi$. The number of hypotheses in the interval 
$(\chi,\chi+\Delta\chi)$ shall be proportional to $\Delta \chi$ and 
independent of $\chi$.  In this way, $\chi$ linearly enumerates the different hypotheses 
we have about the physical situation. Since $\chi$ will have a functional relation to the 
physical parameters of the experiment, $\chi$ can also be thought of as a physical 
parameter. This is how we will use $\chi$ later on. But we will always keep in mind that 
$\chi$ is originally a label enumerating hypotheses. In fact it will turn out that this is all we 
need, and those "actually physical parameters", to which we think $\chi$ to be related, 
will never have to be specified and only serve as a guide to our imagination.

In order to be able to exclude hypotheses by measuring the probability $p$, we need a 
mapping of $\chi$ to $p$. It is important to note that we are completely free in inventing 
this mapping. The reason for this is that the laws existing between the parameters of the 
physical situation and the probability $p$, can be seen as laws between those parameters 
and $\chi$, plus a mapping of $\chi$ to $p$. Figure 3 illustrates the mapping, and how 
we exclude hypotheses by measuring the probability. When $N$ trials have been made, of 
which $L$ resulted in a click at detector 1, we can estimate the true value of $p$ to within 
an uncertainty interval $2 \Delta p$ by means of ordinary probability theory. This permits 
to isolate the corresponding hypotheses as possibly true at the chosen confidence level. As 
can be seen in Fig.3, their labels $\chi$ need not be in just one interval, because we will 
permit different hypotheses to predict the same $p$. And, in order to facilitate mathematical 
manipulation, we will use a mapping where neighbouring values of $\chi$ correspond to 
neighbouring values of $p$. But aside from these technical aspects, what principle criteria 
should we follow?
\begin{figure}
\begin{center}
\epsffile{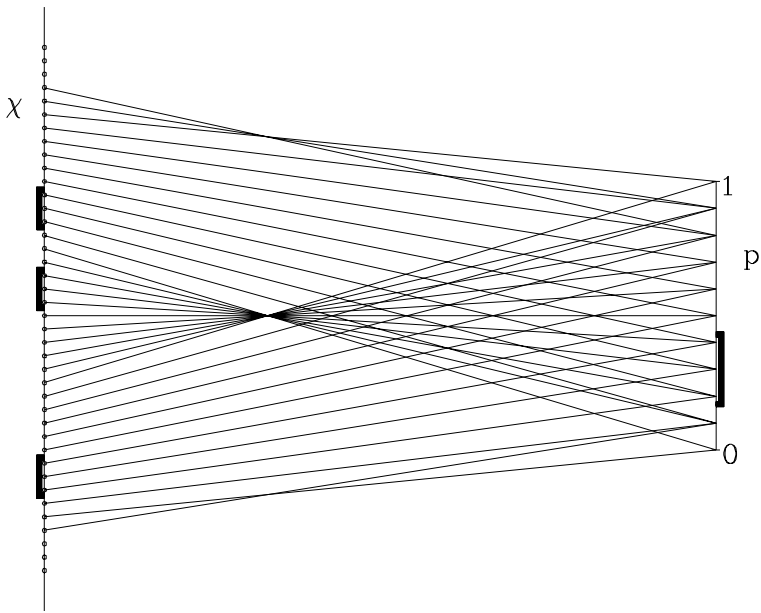}
\caption[Fig.3:]{\small Mapping of physical hypotheses $\chi$ to probabilities 
$p$. Observations permit to determine $p$ to within an interval, which in turn permits to 
narrow down the range of hypotheses, which we can consider as possibly true at the chosen 
confidence level, to the corresponding intervals of $\chi$.}
\end{center}
\end{figure}
 I think there are two properties to be considered. The first is a belief at the bottom of any 
rational endeavour, and in particular of empirical science, which is so elementary that it is 
rarely spelled out:  {\it Through observation our information about the world can only 
increase, never just stay the same, and never decrease}. For our problem this implies that, 
the number of hypotheses still considered as possibly true after a number of observations 
have been made, should strictly decrease with increasing number of observations. The total 
length of the intervals isolated on the $\chi$-axis in Fig.3 should therefore get smaller with 
each additional trial. It is a surprising but well known fact of statistics that almost all well-
behaved mappings of $\chi$ to $p$ {\it do not} have this property.

The second property also concerns the number of hypotheses not yet excluded by the data. 
Consider two runs of the experiment in Fig.2. In each run $N$ trials are made. Let us assume 
detector 1  registered $L$ counts in the first run, and a different number of $L'$ counts in the 
second run. In general, this will isolate different intervals on the $\chi$-axis, whose total 
length will also differ. Therefore, the total number of possibly true hypotheses will also be 
different,  in general, although we have done the same number of observations. This is a 
skewed state of affairs. We would like to be in a position, where $N$ observations always 
isolate equally many hypotheses as possibly true, and where the result $L$ only determines 
the location of these hypotheses on the $\chi$-axis. This would reflect the fact that the 
number of trials, $N$, is not a result. It is something the experimenter can decide.
Therefore, we demand that our mapping of $\chi$ to $p$ has the property that the total 
number of hypotheses not excluded by the experimental data be {\it invariant with respect 
to the outcome of the experiment} and only depend on the number of observations, i.e. trials. 
We may conjecture that this affords great predictive power when we do chains of 
experiments, in which the parameters of one experiment are set according to the results of 
earlier ones, as is the conduct of ordinary science. The space of hypotheses will then be 
multidimensional. But, while the results of the experiments are open, we can know {\it in 
advance} how much of this space will have been excluded by the experiments. We only need 
to fix the number of trials of each of the different probabilistic experiments. I think  once we 
have accepted an intrinsically probabilistic view of the world, it is especially this latter 
property of invariance, which should allow us to define an investigative method which 
nothing can escape. 

The two principle properties we wish to have can be summarized in graphical terms as 
follows: The mapping of the hypotheses $\chi$ to the probabilities $p$ shall be such that, 
while the {\it location} of the isolated intervals on the $\chi$-axis in Fig.3 will depend 
on the result $L$ of the experiment, their {\it total length} must only depend on the 
number $N$ of observations and must strictly decrease with this number. Formally, we want 
to transform the observed random variable $L/N$ to another one, which has the property that 
its standard deviation becomes independent of $L$ when $N$ becomes large.

We will now derive this mapping \cite{Summhammer}.
We have done $N$ trials and registered $L$ clicks in detector 1. The two integers, $N$ and 
$L$, must serve to pin down the fixed, but unknown, probability
 $p_{true}$ that the particle hits detector 1. The random variable $L/N$ is subject to a 
binomial distribution. Since this distribution has a finite standard deviation, it fulfills  
Chebyshev's inequality \cite{Chebyshev}:
\be
Prob\left(\left|p_{true}-L/N\right| >k \sigma\right) \le 
\frac{1}{k^2},
\ee
where the standard deviation $\sigma$ is given by
\be
\sigma = \sqrt{\frac{p_{true}(1-p_{true})}{N}}.
\ee
This inequality means that, the probability that the ratio $L/N$ will deviate from $p_{true}$ 
by more than $k$ standard deviations is less than or equal to $1/k^2$.  The arbitrary 
confidence parameter $k$ only makes sense for $k>1$. Our best estimate for $p_{true}$, 
which we call $p_b$, is
\be
p_b=\frac{L}{N}.
\ee
With this definition, the interval, within which $p_{true}$ can be found with a probability 
larger than $1-1/k^2$, is obtained as
\be
\left( p_b f+\frac{k^2}{2N}f-kf\sqrt{\frac{p_b(1-p_b)}{N} 
+\frac{k^2}{4N^2}}, p_b f+\frac{k^2}{2N}f+kf\sqrt{\frac{p_b (1-
p_b)}{N}+\frac{k^2}{4N^2}}\right),
\ee 
where  $f=(1+\frac{k^2}{N})^{-1}$. For sufficiently large $N$ this reduces to the 
interval
\be
(p_b -k\Delta p_b, \vspace{1mm} p_b+k\Delta p_b)
\ee
with
\be
\Delta p_b = \sqrt{\frac{p_b(1-p_b)}{N}},
\ee
which is the familiar form for the estimate of an unknown probability from observed data.

The width of the uncertainty interval, eq.(5), is $2k\Delta p_b$.  
We will call $\Delta p_b$ the uncertainty of our best estimate $p_b$. The total length of 
the corresponding intervals on the axis of the hypotheses will be proportional to a quantity 
$\Delta \chi$, which we will call the uncertainty of $\chi$, and which is given by
\be
\Delta \chi = \left| \frac {d\chi}{dp_b}\right| \Delta p_b = \left| 
\frac {d\chi}{dp_b}\right|\sqrt{\frac{p_b(1-p_b)}{N}}. 
\ee
(Here we have tacitly assumed the validity of the gaussian approximation of the posterior 
probability distributions both for the estimates of $p$ and of $\chi$, which is justified for 
sufficiently large $N$.)
According to our second principal property, $\Delta \chi$ must only depend on $N$. 
Hence we must have
\be
\left|\frac {d\chi}{dp_b}\right|\sqrt{p_b(1-p_b)}=C
\ee
where $C$ is a positive real constant. From (7) and (8) we obtain, what we had required: 
\be
\Delta \chi = \frac{C}{\sqrt{N}}.
\ee
Because of the absolute sign, integration of (8) yields two solutions:
\be
\chi_{\pm} = \pm C \arcsin(2p-1) + \theta_{\pm}
\ee
where $\theta _{\pm}$ are real constants. However, for any choice of the constants 
$C$, $\theta_+$ and $\theta_-$, the solution $\chi_-$ will be a linear function of 
the solution $\chi_+$, such that it is actually not a different mapping of hypotheses to 
probabilities. Therefore, we only need to consider the solution $\chi_+$, which we will 
simply call $\chi$, and the arbitrary constants will be $C$ and $\theta$.
The inverse of eq.(10) is thus
\be
p= \cos^2\left(\frac{\chi - \theta}{2C}\right).
\ee
It is depicted graphically in Fig. 4. An experimentally established uncertainty interval of $p$ 
defines an infinite number of corresponding uncertainty intervals on the axis of $\chi$, 
whose width is invariant with respect to the outcome of the experiment. 
Note that we can infer numerical values to label the hypotheses to which the result of a 
concrete experiment points, only after fixing $C$ and $\theta$ .

Although eq.(11) is common knowledge in statistics, its significance to physics seems to have 
gone unnoticed until the work of Wootters on distinguishability in quantum experiments 
\cite{Wootters}. 

\begin{figure}[ht]
\begin{center}
\epsffile{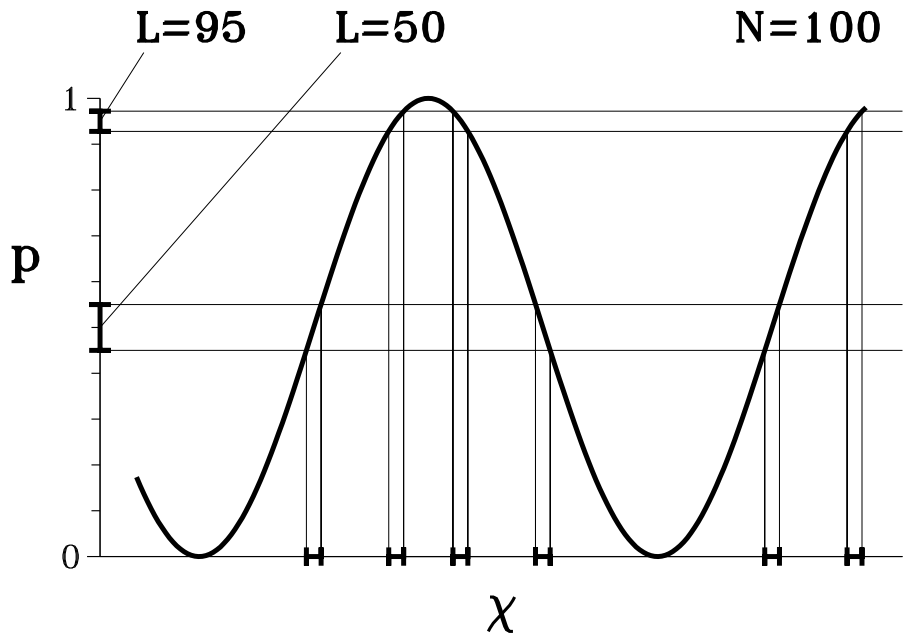}
\caption[Fig.4:]{\small Mapping from observed probability, $p$, to the label of 
hypotheses, $\chi$, for two experiments with the same number of trials, $N=100$. In one 
experiment detector 1 clicked $L=95$ times, in the other only $L=50$ clicks occurred. The 
uncertainty ranges on the $p$-axis are very different, but on the $\chi$-axis they are equal. 
Therefore, both experiments excluded equally many hypotheses.}
\end{center}
\end{figure}

As we have mentioned before, and as is evident from Fig.4, the invariance property of 
$\Delta \chi$ permits to make a very powerful statement. One can specify with the 
confidence level fixed by the parameter $k$ of eq.(1), the fraction of hypotheses excluded by 
$N$ trials of the experiment.  The excluded fraction is $1-\frac{2k}{\pi 
\sqrt{N}}$. This implies for instance that, if we just {\it plan} to do 1000 trials, we 
immediately know we have excluded 94.0\% of the hypotheses with a confidence level 
corresponding to $k=3$. Such concrete {\it advance knowledge} is impossible with any 
other mapping of hypotheses to probabilities. In order to see this, suppose we had, instead of 
eq.(11), a linear mapping of hypotheses to probabilities. Following eq.(5), the excluded 
fraction of hypotheses would be given by $1-2k\Delta p$, which, for $N=1000$ and 
$k=3$, can be anywhere between 90.5\% and 99.1\% (eq.(4) has been used for the limits of 
$p$ at 0 and 1, because the approximate form of eqs.(5) and (6) is invalid there). But without 
actually {\it doing the experiment and obtaining the data} we could not say anything more 
specific.

Another graphical presentation of the mapping of $\chi$ to $p$ is shown in Fig.5a. The 
circle is a segment of the axis of $\chi$ which corresponds to one full period. The axis of 
the probability $p$ is the diameter of the circle. Projecting the uncertainty limits of $p$ onto 
the circle gives the corresponding uncertainty limits of $\chi$. Yet another representation 
of $\chi$ is obtained by placing this drawing into the complex plane such that the bottom 
of the circle is at the origin, and then tilting it by an arbitrary angle, as shown in Fig.5b. This 
representation is used by quantum theory. Later we will see that it is more powerful for 
forming predictions than the purely real representations.
\begin{figure}
\begin{center}
\epsffile{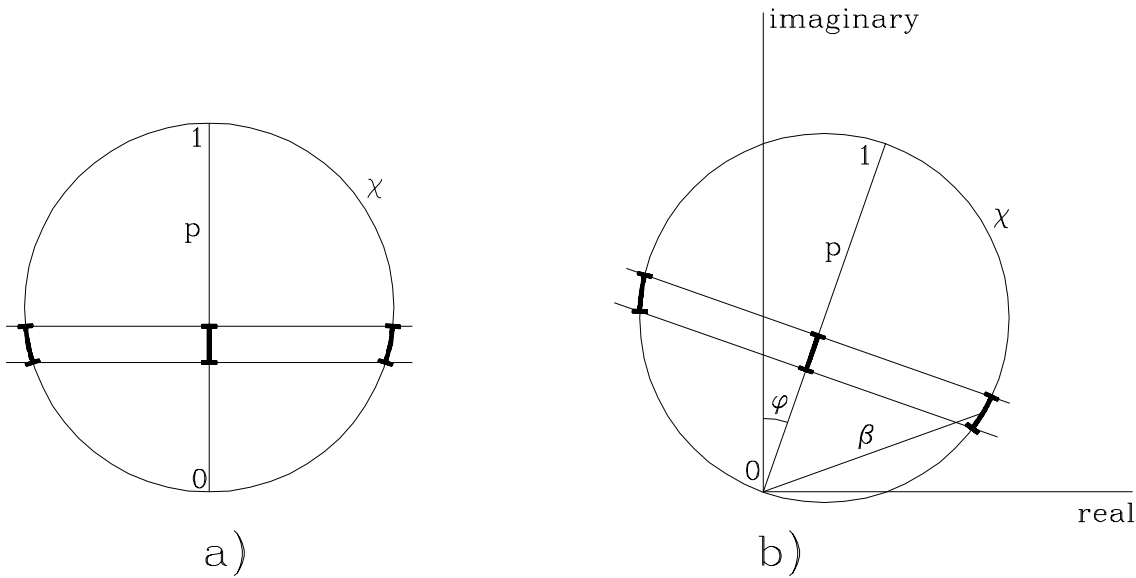}
\caption[Fig.5:]{\small {\bf a)} The mapping of eq.(11) depicted as projecting 
points of $p$-axis  perpendicularly outwards onto a circle, whose diameter is the $p$-axis.
{\bf b)} Embedding this into the complex plane, such that the origin coincides with the 
point $p=0$ defines a mapping of probability $p$ to a point $\beta$, which lies on the 
circle that represents the $\chi$-axis. One finds $p=|\beta|^2$. The phase of 
$\beta$ is free, because it depends on the arbitrary choice of the tilt angle $\varphi$ 
between the $p$-axis and the imaginary axis.}
\end{center}
\end{figure}

To summarize this section, we found a unique mapping of hypotheses to probabilities. From 
the point of view of probability theory we simply derived that random variable, whose 
standard deviation asymptotically becomes an invariant under different outcomes $L$ of the 
experiment. That is why our finding has nothing to do with elementary physical experiments, 
in particular. (A similar observation was made by Brukner and Zeilinger, when trying to find an  
invariant information measure for 2-level systems \cite{Caslav}.) However, we will see 
that there is a deeper reason why $\chi$ or $\beta$ are useful quantities in binomial 
experiments like elementary quantum yes-no setups, and less useful ones in seemingly 
analogous everyday procedures like estimating the fraction of red cars in a city by 
determining their fraction in just a few streets. 

In the following sections we will derive other quantities from more complex experimental 
schemes. We will from now on no longer emphasize that by means of an experiment we 
eliminate hypotheses. We will switch to the usual manner of saying that we measure 
quantities. And we will reserve the term "physical quantity", to measurable quantities  which 
have the same invariance properties as we required from our quantity $\chi$: The 
uncertainty interval must get smaller with each additional observation, and it must be 
invariant under different outcomes of the experiment. 

\section{Observation of a probability as a function of a parameter}

We will now measure the probability of a click in detector 1 in the experiment of Fig.2 for 
several different physical situations, which we will characterize by the preparation parameter 
$t$.  (This need not be time.) We would determine $t$ by means of auxiliary measurements, 
which would ultimately have to be resolvable into observations of probabilities, but would, of 
course, not involve detections at either detector 1 or detector 2. For the present it will suffice 
to assume that we can set $t$ and verify this setting. Let us measure the fraction of clicks at 
detector 1 as a function of a {\it sequence} of different settings $t_j$, $j=1,...,M$. For 
simplicity we assume the values of the $t_j$ to be aequidistant, and without restricting 
generality we can let $t$ have integer values , thus $t_j =j$. But we assume that arbitrary 
values in between can also be set. With each setting $t_j$ we do $N_j$ trials. We will first 
analyse the experiment by means of the the real label $\chi$ and then repeat the analysis 
by using the complex label $\beta$.

\subsection{Real labels for hypotheses}

Following eq.(10) we derive the physical quantities $\chi_j$.  In accordance with eq.(9) 
we obtain their respective uncertainties as
\be
 \Delta \chi_j = N_j^{-1/2}
\ee
where we set $C=1$. 
Having done this, what other quantities can we form, whose uncertainties have the same 
invariance properties as the $\Delta \chi_j$ such that we can consider them as {\it 
physical quantities}? Clearly, such other quantities can only be linear functions of the 
$\chi_j$. 
Let them be denoted by $g_l$, $l=1,...,M$:
\be
g_l = \frac{1}{M}\sum_{j=1}^M a_{lj}\chi_j +\eta_l,
\ee
where $a_{lj}$, and $\eta_l$ are arbitrary real constants. In order that at each setting 
$t_j$ additional data will contribute equally to a decrease of uncertainty of the $g_l$, we 
must require $a_{lj}=\pm 1$. The uncertainty intervals are then
\be
\Delta g_l = \Delta g =\frac{1}{M}\sqrt{\sum_{j=1}^M 
\frac{1}{N_j}}.
\ee
Depending on which of the $a_{lj}$ we set $+1$ and which we set $-1$ the quantities 
$g_l$ may be more or less meaningful. 
But we can obtain a particularly useful linear combination when we 
give up the constraint on the $a_{lj}$ and let them be coefficients in a Fourier series.
For simplicity we assume that $M$ is odd, thus we write $M=2K+1$, and we divide the 
$g_l$ into the sine coefficients $s_l$ and the cosine coefficients $c_l$. Then, for 
$l=0,...,K$: 
\be
c_l=\frac{1}{M}\sum_{j=1}^M \chi_j \cos\left(\frac{2\pi j 
l}{M}\right) +\eta_l^{(c)}
\ee
and
\be
s_l=\frac{1}{M}\sum_{j=1}^M \chi_j \sin\left(\frac{2\pi j 
l}{M}\right) + \eta_l^{(s)},
\ee
where we have $s_0=0$, so that there are again just $M$ different quantities. As before, 
we also have arbitrary real additive constants, $\eta_l^{(c)}$ and $\eta_l^{(s)}$. 
The uncertainty intervals are
\be
\Delta c_l=\frac{1}{M}\sqrt{\sum_{j=1}^M 
\frac{\cos^2\left(\frac{2\pi j l} {M}\right)}{N_j}}
\ee
and
\be
\Delta s_l=\frac{1}{M}\sqrt{\sum_{j=1}^M 
\frac{\sin^2\left(\frac{2\pi j l}
{M}\right)}{N_j}}.
\ee
However, here we are faced with the problem that the cosine function becomes zero for 
certain combinations $jl$, for which obtaining additional data with the setting $t_j$ does not 
result in a decrease of $\Delta c_l$. The same holds for other combinations $jl$ for the 
sine function and the corresponding $\Delta s_l$. Therefore, according to our definition of 
a physical quantity, not all of the $c_l$ and $s_l$ would qualify for this status. 

In order to remedy the situation we are led to consider assigning to {\it one} physical 
quantity {\it two} numbers, instead of just one. We will do this by expanding into the 
plane of complex numbers. We can now define for $l=-K,...,K$:
\be
h_l=c_l+i s_l = \frac{1}{M}\sum_{j=1}^M \chi_j \exp\left( i\frac{2\pi 
jl}{M}\right) + \zeta_l,
\ee
where $\zeta_l=\eta_l^{(c)}+i\eta_l^{(s)}$ are arbitrary complex constants.
If the $h_l$ are to be accepted as physical quantities, their uncertainties must only depend 
on the numbers of trials $N_j$ made at the various settings $t_j$, and must strictly decrease 
when {\it any} of the $N_j$ increases. Since our definition of uncertainty is equal to the 
standard deviation of the respective random variable when the number of trials becomes 
large, we can again employ the gaussian approximation and obtain
\be
\Delta h_l = \sqrt{\sum_{j=1}^{M} \left| \frac{\partial 
h_l}{\partial \chi_j}\right|^2 \left(\Delta \chi_j\right)^2 } = 
\frac{1}{M}\sqrt{\sum_{j=1}^M \frac{1}{N_j} }.
\ee
Obviously, the $\Delta h_l$ have the required properties, so that the $h_l$ are {\it 
physical quantities}. Moreover, all the $h_l$ have the same uncertainty.

Nevertheless, there is a conspicuous asymmetry between the physical quantities $\chi_j$ 
and the equally physical quantities $h_l$. The former are by design real, while the latter are 
in general complex. It is therefore suggestive to start right away with a complex 
representation of $\chi$ as is shown in Fig.5b. We will do so, but we do not want to 
motivate this step by formal elegance. Rather, we will first show that {\it predictions} 
based on the quantitities $h_l$ as derived from real $\chi_j$ are too consistent to be 
credible.

\vspace{3mm}
{\bf PREDICTIONS BASED ON $\chi$:} From a formal point of view a prediction is no 
different to a measurement. It is the calculation of just another random variable from observed 
data. But, of course, a prediction contains the assumption that from data known to the 
observer something can be said about the range of possible values of random variables that 
the observer will obtain from {\it other} measurements. According to what rules should 
such predictions be made? Normally, models of the world, or at least of a particular aspect of 
the world are invented for the purpose. We shall refrain from this. In the spirit of this paper a 
prediction must become more accurate with an increase of the number of trials of the 
probabilistic experiments from whose outcomes the prediction is calculated. And, of course, 
the uncertainty of the prediction must be an {\it invariant} of the outcomes of these 
experiments. This will ensure the predicted quantity to be a physical quantity. To form such 
quantities, we can again rely on a linear combination of observed physical quantities.

Presently we will use the observed $h_l$ to form predictions of the values of $\chi$ that 
will be observed for settings of our control parameter $t$, for which we have not yet made 
an experiment. Naturally, we have no guarantee that our predictions will hold true when 
testing them. Let the predictions be denoted by $\chi ' (t)$:
\be
\chi ' (t) = \sum_{l=-K}^{K} \left( h_l - \zeta_l \right)\exp{\left(-
i\frac{2 \pi l}{M} t\right)} = \frac{1}{M} \sum_{l=-K}^{K} 
\sum_{j=1}^M \chi_j \exp{\left[i \frac{2 \pi l}{M}(j-t)\right]}.
\ee
Of course, when $t$ is an integer $j$ from 1 to M we have $\chi'(t=j)=\chi_j$. The 
uncertainty of $\chi'(t)$ due to an uncertainty of the observed quantity $\chi_j$ is given 
by
\be
\Delta_j \chi'(t) = \left|\frac{\partial \chi'(t)}{\partial 
\chi_j}\right|\Delta \chi_j=\frac{1}{M}\left|1 + \sum_{l=1}^{K} 2 
\cos{\left[\frac{2 \pi l}{M}(j-
t)\right]}\right|\frac{1}{\sqrt{N_j}}.
\ee
As expected, we have $\Delta_j \chi'(t=j)=\Delta\chi_j$. If $t$ is not an integer, 
$\Delta_j \chi'(t)$ depends most strongly on the uncertainty of the nearest data point, 
and less strongly on the uncertainties of remote data points, as shown in Fig.6. This meets our 
intuition as to which data points are the most important for making the prediction. A closer 
inspection of eq.(22) and of Fig.6 reveals that for $t$ not an integer we have $\Delta_j 
\chi'(t)\ne 0$ for all $j$, $j$=1,...,M. This means that {\it all} observed data enter into 
the prediction $\chi'(t)$. Most importantly, however, $\Delta_j \chi'(t)$ is {\it 
invariant} of the particular data that we observed, and it strictly decreases with an increase 
of the number of observations, $N_j$. The overall uncertainty of the prediction is
\be
\Delta \chi'(t) = \sqrt{\sum_{j=1}^M \left| \Delta_j 
\chi'(t)\right|^2},
\ee
and it, too, is invariant of the outcomes of the experiments and decreases with an increase of 
any of the $N_j$. This qualifies the prediction $\chi'(t)$ as a physical quantity.
\begin{figure}[ht]
\begin{center}
\epsffile{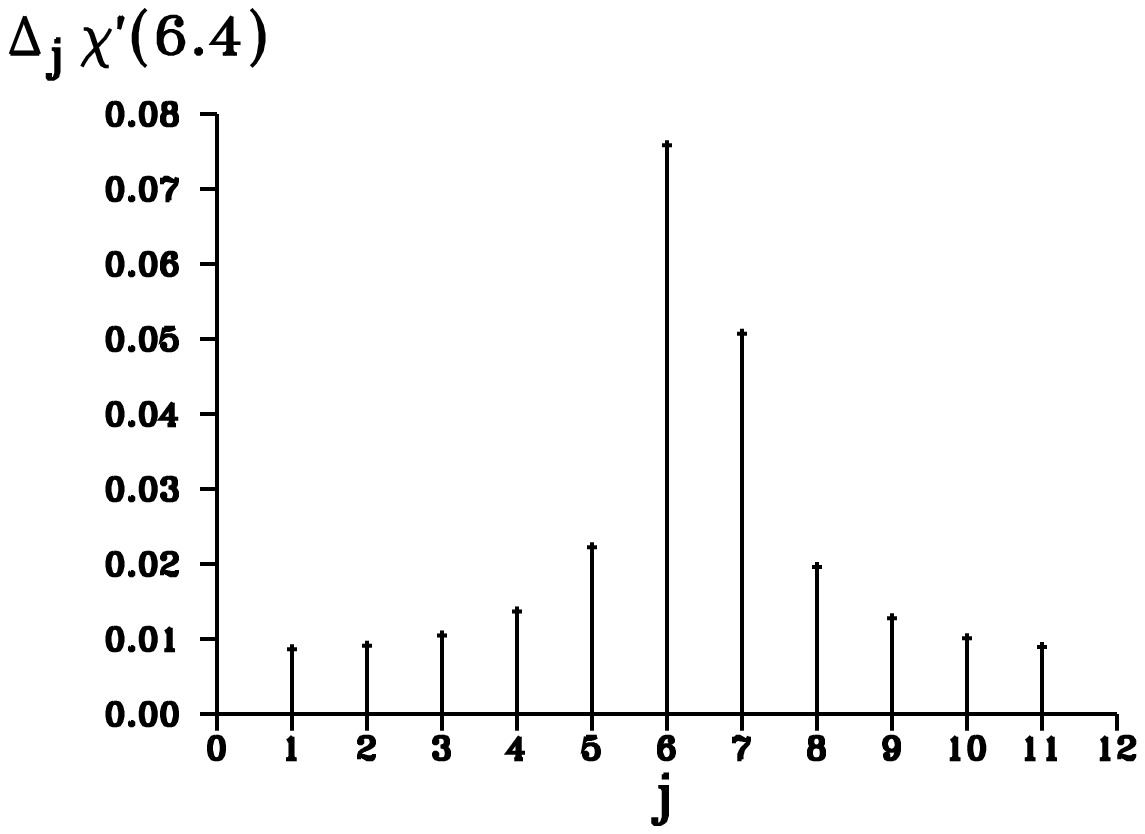}
\caption[Fig.6:]{\small Uncertainties of the prediction $\chi ' (t)$ with $t=6.4$ as 
a function of the uncertainties of the individual data points $\chi_j$, j=1,...,M, in an 
example with M=11 and $N_j = 100$ trials for all $j$. The uncertainty of the closest data 
point ($j=6$) has the strongest influence on the uncertainty of the prediction. Data need not 
be known to calculate the {\it uncertainty} of the prediction.}
\end{center}
\end{figure}

In order to test the prediction we will have to calculate $p[\chi'(t)]$ for the desired $t$ 
using eq.(11). We must, of course, decide on the constants $C$ and $\theta$ already 
when determining the values of $\chi_j$ from the observed probabilities. Then we can do 
an experiment with the setting $t$ and compare this new observation with the prediction.

Unfortunately, we cannot trust the prediction $\chi'(t)$. First, we note an ambiguity in 
assigning a value to $\chi$ from a measured probability $p$ using eq.(11). It is evident 
from Fig.4 that, fixing the constants $C$ and $\theta$ is not sufficient to obtain a unique 
value for $\chi$, because of the sinusoidally periodic relationship between $p$ and 
$\chi$. Even if we select a specific period, there are always two possible values for 
$\chi$, except when $p=0$ or $p=1$. So, in general, a series of M different settings of 
$t$ permits to derive $2^M$ different sets of the physical quantities $h_l$, $l$=-K,...,K, 
and hence equally many different predictions $\chi'(t)$ for any given $t$. Two examples 
are shown in Fig.7.
\begin{figure}[ht]
\begin{center}
\epsffile{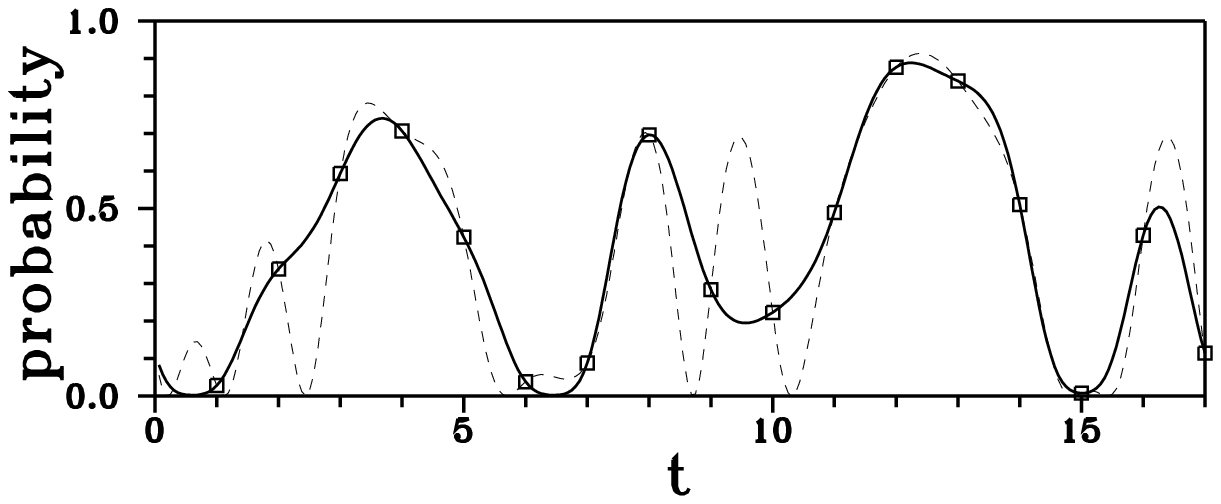}
\caption[Fig.7:]{\small The squares show the best estimates of measured 
probabilities for M=17 different settings of the parameter $t$. The full line is a prediction 
made by inferring values of $\chi$ in the interval [0,$\pi$] from the observed 
probabilities and then applying eq.(21) to obtain the predicted $\chi'$. The dashed line is 
another prediction, where the values of $\chi$ were inferred by randomly projecting the 
observed $p$ to the left or right side of the circle of Fig.5a, so that $\chi$ could have 
values in the interval [-$\pi$,$\pi$].}
\end{center}
\end{figure}

The possibility of so many predicted functions is clearly unsatisfactory.
Not only do we predict a range of different experimental outcomes for any new setting of the 
control parameter $t$, but our method convinces us that each of them is consistent with the 
known data. We can wonder why the $M$ observed probabilities should be due to only $M$ 
underlying physical quantities $h_l$? There could be many more $h_l$, $|l|>K$, that 
contribute to the coming about of the data. But our method gives us no hint about their 
existence, although it does not exclude that we invent some and see whether this is 
compatible with the data. Therefore, what we would like to have is a means which can tell 
us on the basis of the {\it available data}, that there might be further quantities $h_l$, 
$|l| > K$, because the predictions derived from the data lead to inconsistencies. For this 
purpose we shall investigate the complex representation of $\chi$ as shown in Fig.5b.

\subsection{Complex labels for hypotheses}

From Fig.5b one obtains by simple trigonometry the complex representation of $\chi$  as
\be
\beta = \left( \sqrt{p(1-p)} \pm ip\right) e^{-i\varphi} .
\ee
Assigning a numerical value to $\beta$ from experimental data requires the choice of a 
phase $\varphi$ and a decision whether $p$ shall be projected to the left or the right half 
of the circle. 

The uncertainty of $\beta$ is given by
\be
\Delta \beta = \left|\frac{d\beta}{d p}\right|\Delta p = 
\frac{1}{2 \sqrt{N}}.
\ee
It is invariant under different outcomes of the experiment. Therefore, $\beta$ is a physical 
quantity according to our definition. Because $\beta$ is such an important quantity --- we 
know it in quantum theory as the probability amplitude for the specific outcome --- we shall 
also show that the standard deviation does indeed become independent of the experimental 
outcome when the number of trials is large.

The {\it random variable} $\beta$ is given by
\be
\beta (L/N) = \sqrt{\frac{L}{N}\left(1-\frac{L}{N}\right)} + i 
\frac{L}{N}.
\ee
Here, $L$ is again the number of outcomes "1" in $N$ trials of the experiment of Fig.2. The 
probability for "1" in a single trial is given by $p$.  We neglect a general phase in $\beta$, 
because this would only amount to a rotation around the origin in the complex plane. 
Similarly, it is sufficient to consider only one projection of $p$ onto the circle (see Fig.5b). 

The expectation value of $\beta$ is given by summing over all possible experimental 
results with the weights according to the binomial distribution
\be
\bar{\beta} =\sum_{L=0} ^{N} \frac{N!}{L!(N-L)!}p^L (1-p)^{N-L} 
\beta(L/N).
\ee
The dispersion of $\beta$ is given by $\sigma^2$, where $\sigma$ is the standard 
deviation. $\sigma^2$ is defined as the expectation value of the absolute square of the 
difference between $\beta$ and $\bar{\beta}$
\be
\sigma^2(\beta) =\sum_{L=0} ^{N} \frac{N!}{L!(N-L)!}p^L (1-p)^{N-L} 
\left|\beta(L/N)-\bar{\beta}\right|^2.
\ee
Elementary transformation yields
\be
\sigma^2(\beta)= \left|\bar{\beta}\right|^2 + p + i p(\bar{\beta}-
\bar{\beta^*}) - (\bar{\beta}+\bar{\beta^*})\sum_{L=0} ^{N} 
\frac{N!}{L!(N-L)!}p^L (1-p)^{N-L} \sqrt{\frac{L}{N}\left(1-
\frac{L}{N}\right)}.
\ee
Noting that the imaginary part of $\bar{\beta}$ is equal to $p$, and that 
\be
\left|\bar{\beta}\right|^2 = \left[Re(\bar{\beta})\right]^2 + 
\left[Im(\bar{\beta})^2\right]^2
\ee
we can simplify further to
\be
\sigma^2(\beta)=-\left[Re(\bar{\beta})\right]^2 + p(1-p).
\ee
Since the first term on the right side, i.e. the real part of the expectation value of $\beta$, 
is difficult to evaluate analytically it was obtained numerically. The results for $\sigma$ 
are shown in Fig.8. One notes that $\sigma$ does indeed become independent of $p$ and 
that it approaches the uncertainty of $\beta$ given in eq.(25): $\sigma \rightarrow 
\Delta\beta = 1/(2\sqrt{N})$.
\begin{figure}[ht]
\begin{center}
\epsffile{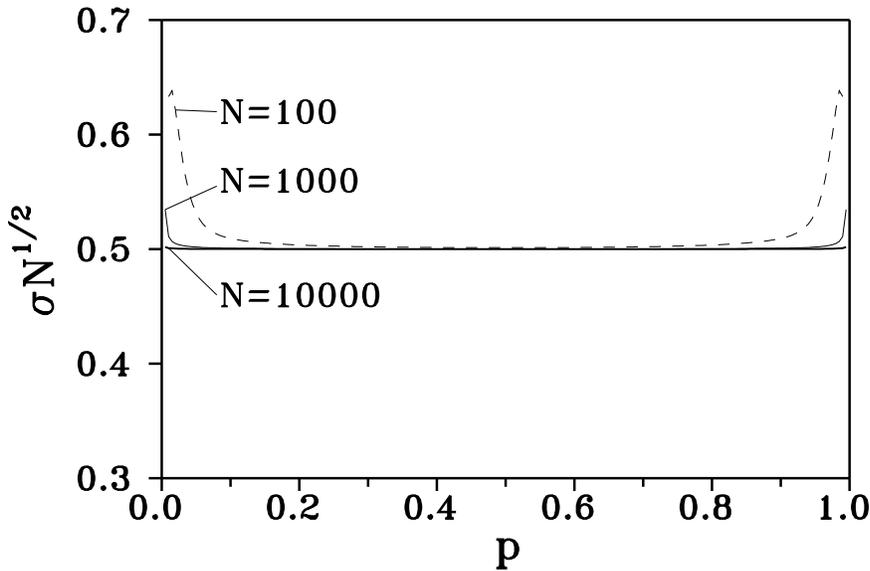}
\caption[Fig.8:]{\small The standard deviation $\sigma$ of the random variable 
$\beta = \sqrt{(L/N)(1-L/N)} + i L/N$ as a function of the intrinsic probability $p$ to 
obtain the outcome "1" in the experiment of Fig.2. As the number of trials $N$ increases 
$\sigma$ becomes independent of $p$. In the plot $\sigma$ was multiplied by 
$\sqrt{N}$ to eleminate the overall decrease of the standard deviation with increasing 
$N$. }
\end{center}
\end{figure}

\vspace{3mm}
Let us now look at the observation of a probability as a function of an experimental parameter 
$t$, as we did before. Again we assume $M$ different settings equidistant in $t$. Also, we 
let $M$ be odd to have the formal advantage of summation indices symmetric around 0 later 
on. At each setting we do $N$ trials of the experiment and get the outcomes 
$L_1$,...,$L_M$. Then we calculate the best estimates for the $\beta_j$, $j=1,...,M$ as
\be
\beta_j = \left( \sqrt{\frac{L_j}{N}\left(1-\frac{L_j}{N}\right)} + 
i\frac{L_j}{N}\right) e^{-i\varphi_j}.
\ee
Here, we projected $p$ onto the right half of the circle, but the choices of $\varphi_j$ 
are still free. Now we form linear combinations of the various $\beta_j$ which correspond 
to the Fourier expansion:
\be
v_l = \frac{1}{\sqrt{M}}\sum_{j=1}^{M}\beta_j \exp{\left(i\frac{2 
\pi}{M}jl\right)}
\ee
for $l=-K,...,K$ and $2K+1=M$. We have introduced the scale factor $1/\sqrt{M}$ with 
the purpose to obtain equal uncertainties of the $v_l$ and the $\beta_j$. By virtue of the 
fact that the dispersion of a sum of random variables is equal to the sum of the dispersions of 
these random variables the uncertainties $\Delta v_l$ will also be independent of the 
outcomes $L_1,...,L_M$. We have:
\be
\Delta v_l =\sqrt{\sum_{j=1}^{M} \left|\frac{\partial 
v_l}{\partial \beta_j}\right|^2 \left(\Delta 
\beta_j\right)^2}=\frac{1}{2\sqrt{N}},
\ee
where we have made use of eq.(25). The invariance of the uncertainties of the $v_l$ under 
different experimental outcomes, and the strict decrease of the uncertainties with more trials, 
qualifies the $v_l$ as physical quantities. Clearly, this is also true if the $M$ experiments 
were not all done with the same number of trials and eq.(34) would not reduce to such a 
compact form. With this, we can now turn to forming predictions.

\vspace{3mm}
{\bf PREDICTIONS BASED ON $\beta$:} Using the quantities $v_l$ we form predictions 
for arbitrary settings of the parameter $t$ by means of Fourier synthesis
\be
\beta'(t) = \frac{1}{\sqrt{M}}\sum_{l=-K}^K v_l^* \exp{\left(-
i\frac{2\pi}{M}lt\right)}=\frac{1}{M}\sum_{l=-K}^K \sum_{j=1}^M 
\beta_j \exp{\left[i\frac{2\pi}{M}l(j-t)\right]}.
\ee
The uncertainty of $\beta'(t)$ for a given experimental setting $t$ is
\be
\Delta\beta' = \sqrt{\sum_{j=1}^M 
\left|\frac{\partial\beta'}{\partial\beta_j}\right|^2(\Delta
\beta_j)^2}=\sqrt{\frac{1}{M}\sum_{j=1}^M 
(\Delta\beta_j)^2}=\frac{1}{2\sqrt{N}}.
\ee
It is thus also independent of the experimental outcomes and decreases with the number of 
trials. Therefore, the prediction $\beta'(t)$ is a physical quantity. Note that $\Delta 
\beta'$ is also independent of $t$.

Now we want to compare predictions made on the basis of $\chi$ to those made on the 
basis of $\beta$. For this purpose we have used the same data as in Fig.7 and calculated 
predictions $\beta'(t)$, which are shown in Fig.9. Clearly, we have an infinite number of 
possibilities to form the prediction function $\beta'(t)$, because to each observed 
$\beta_j$ we can assign an arbitrary phase $\varphi_j$. But here, Fig.9 points to a 
severe problem: The transformation back to a probability according to $p=|\beta'|^2$ 
can result in values $p>1$, which are clearly meaningless.
\begin{figure}[ht]
\begin{center}
\epsffile{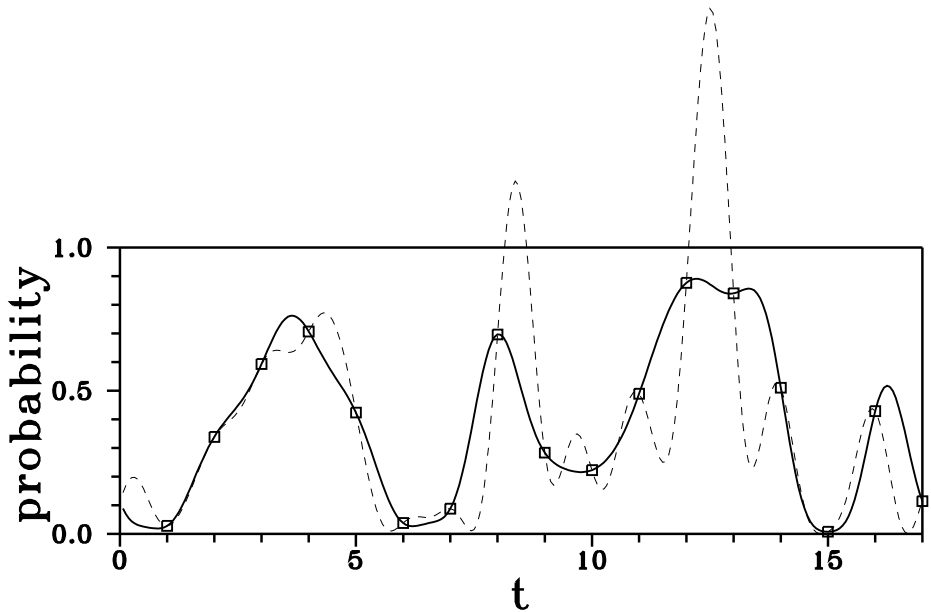}
\caption[Fig.9:]{\small The squares show the best estimates of the same measured 
probabilities as in Fig.7 for the different settings of the parameter $t$. The full line is a 
prediction made by inferring values of $\beta_j$ with the free phases $\varphi_j$ all 
set to zero and then applying eq.(35) to obtain the prediction function $\beta'(t)$.  The 
dashed line is an example of the prediction function when random phases $\varphi_j$ are 
assigned to the inferred values of $\beta_j$.}
\end{center}
\end{figure}

The reason for the occurrence of nonsensical predictions is that the inferred $v_l$ do {\it 
not} fulfill the relation
\be
\sum_{j=1}^M \left|v_l\right|\le 1,
\ee
such that the summation (35) is not bounded by 1. But, of course, it is possible that there are 
observations in which the data fulfill this relation. This tells us something: 
The arbitrary phases $\varphi_j$ we assigned to each observed $\beta_j$ cannot be 
chosen as freely as we thought. In fact, we are forced to consider them determined by the 
experimental conditions, at least relative to each other, because the predictions for the 
probability for not yet observed settings $t$ depend crucially on them. We will remark on the 
implications in the discussion. 

The experiment can set limits on the possible combinations of the various $\varphi_j$, 
such that the above inequality is fulfilled.
However, it could well happen, that no combination of the assigned phases $\varphi_j$ 
leads to a prediction function which is bounded by $\left|\beta'(t)\right|\le1$.
Therefore we can use the above inequality to separate the kinds of parameters, as a function 
of which we are observing a probability, into two groups. Those, which fulfill this inequality, 
and those which don't.

\noindent
This distinction is not new. A similar distinction is quite natural in quantum physics, where 
we can observe a probability as a function of a parameter, and where the sum of these 
probabilities may or may not have to be limited by 1. Consider the following three examples:

\noindent
{\bf i)} Observation of the arrival of a particle in a region of space. Our detector "1" would 
cover this region, detector "2" would represent the rest of space. The parameter $t$ would 
label adjacent {\it regions} in space to which we move detector "1" successively and 
repeat the experiment with the identically prepared particle. Clearly, the sum of the 
probabilities at detector "1" cannot exceed 1.

\noindent
{\bf ii)} Quite generally, measurement of the probability to find the particle in one of 
several possible states. Our detector "1" would detect for different states as a function of the 
parameter $t$, such that each new setting of $t$ would test for another possible state. 
Detector "2" would in each case collect together all the other states. Again, the sum of the 
probabilities measured at detector "1" cannot go beyond 1.

\noindent
{\bf iii)} Measurement of the probability to find a spin-1/2 particle in the state $|+z>$, 
when it has been prepared in the state $|+b>$ .  Detector "1" would record the state 
$|+z>$, and detector "2" would record $|-z>$. Our parameter $t$ would label the angle 
between $\vec b$ and $\vec z$. Since we can adjust the successive settings of the angle 
arbitrarily close, summation of the measured probabilities of arrival in detector "1" has no 
upper limit.

\vspace{3mm}
\noindent
These examples show that the observation of a probability as a function of a parameter has 
something to do with the concept of "system". So far we have avoided speaking of "systems", 
because we would have introduced an unjustified ad hoc structure. But now we have a first 
indication that it may be meaningful to start categorizing data as pertaining to "systems", 
although in this paper we cannot penetrate further into this question. 
It is necessary to emphasize that working with the purely real quantities $\chi_j$ would 
never have pointed us to the problem. There, the prediction function is always well behaved, 
because the predicted $\chi'$ can have any value and the derived expectation for the 
probability will never leave the interval [0,1]. Using the $\beta_j$ is thus a more 
discriminatory method. The complex representation of observed probabilities by means of 
$\beta$ extracts potentially more information from the data than the real representation 
$\chi$. 
Therefore, we will from now on use only $\beta$. As we have already remarked, it is what 
quantum theory calls the {\it probability amplitude}.

At this point we must mention that the existence of a deeper structural relation between 
probability theory and the quantum theoretical calculus has already been emphasized by 
Lande \cite{Lande}, and is also shown by Peres \cite{Peres}. A more mathematical 
work of Fivel comes to the same conclusion \cite{Fivel}. And that the superposition 
principle is not of "physical" origin was recently elucidated by Orlov \cite{Orlov}.

\section{Correlations}

\vspace{5mm}
So far we have assumed that our probabilistic experiment of Fig.2 is performed under well 
defined conditions, which are themselves ultimately verified by a huge number of elementary 
detectors. At these exterior detectors we only verify the {\it probabilities} of clicks. If 
these are the same as in the previous trials, at the chosen level of accuracy, we say we have 
the same conditions. In actual experiments one rarely checks such probabilities themselves. 
Instead one reads quantities, which are at least in principle derived from them, from 
indicators of instruments, e.g. readings of lengths, currents, etc. (We imagined our parameter 
$t$ to be verified in this manner.) And in large part the check of the conditions are 
evaluations of the human senses, which inform us only about mean values, anyway (e.g. 
when the experimenter verifies by tactile and visual inspection that all the cables are 
connected in the same way as in the previous trial). Therefore, the data we subsumed under 
the heading {\it conditions}, are not analysed at the level of clicks. 

We will now improve on this situation by looking at one of the conditions in detail. For 
simplicity we will pick out one condition, which up to now was only monitored as the relative 
frequency with which a click occurred in one of two detectors.  Now, occurrence of the click 
in one detector shall mean another condition than occurrence of the click in the other, so that 
we are doing the observation at our original two detectors under {\it two different 
conditions} (we exclude other combinations of clicks at the two new detectors as possible 
conditions). The setup is shown in Fig.10. It is the familiar arrangement of experiments on the 
Einstein Podolsky Rosen paradoxon. 
\begin{figure}[ht]
\begin{center}
\epsffile{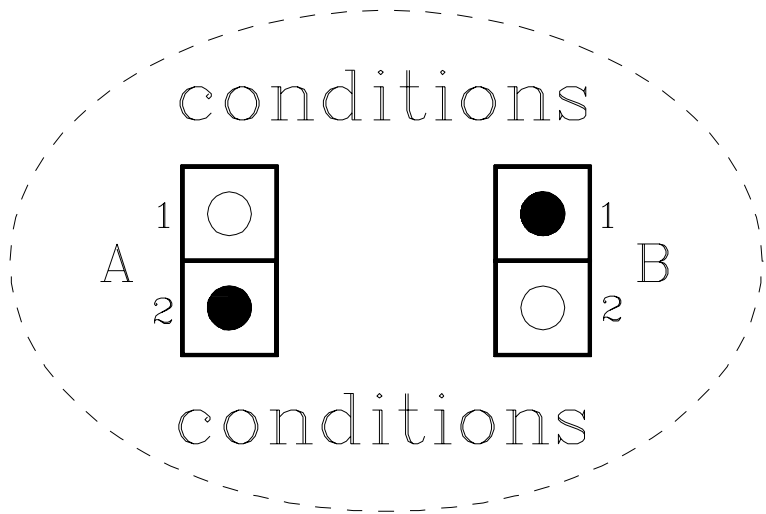}
\caption[Fig.10:]{\small Two yes-no experiments. At each site a click can happen in 
one of two detectors. The physical conditions determine the probabilities for the four possible 
outcomes.}
\end{center}
\end{figure}
Observations are made at two sites, each equipped with two detectors. We observe 
coincidences of clicks of one detector at site A with one at site B, under well defined 
conditions. (Coincidences need not be restricted to a specific time interval. Only some 
procedure of relating data of site A with those at site B is needed, which is independent of 
the data itself.) We do not regard clicks in both detectors at just one site, nor more than one 
click in a detector. Therefore we have four possible events: ($A_1$,$B_1$), ($A_1$,$B_2$), 
($A_2$,$B_1$), ($A_2$,$B_2$). From the point of view of the observer at site A, he or she 
is observing the relative frequencies of clicks in the detectors under the two different 
conditions that at site B either detector 1, or detector 2 fires. Conversely, the observer at site 
B registers the relative frequency of clicks in the two detectors under the two different 
conditions that at site A either detector 1, or detector 2 fires. The overall view is that we have 
a probabilistic experiment under well defined conditions, where one trial can have four 
possible outcomes.

Let us first assume the point of view of the observer at one site, say A. Observer A may start 
out by neglecting the clicks at site B, i.e., he only convinces himself that the relative 
frequency of clicks at $B_1$ and $B_2$ remains constant, just as all other conditions, while 
he is doing the $N$ trials. He is then effectively doing the sort of experiment of Fig.2. (In 
practice A may not even be aware of the existence of site B. Then he just has to {\it 
assume} that all unknown conditions are constant. In fact physics relies on this assumption, 
because an experimenter cannot monitor the whole world while doing some particular 
observation.)

In $N$ trials detector $A_1$ fires $L_A$ times, detector $A_2$ fires $N-L_A$ times. From 
these data observer A can extract the probability $p_A$ to obtain a click at $A_1$, and the 
corresponding physical quantity $\beta_A$ in the manner discussed in section 4.2, and he 
may measure these quantities as a function of various parameters. It may well happen that he 
finds no parameters by which he can adjust $p_A$ in its full range from 0 to 1. In that case 
observer A can come to two different conclusions:

\noindent
i) He may assume that this specific probabilistic experiment simply cannot yield all of the 
logically conceivable results. The most general conclusion  would be that the present state of 
the world excludes some possibilities. ("Constants" of nature would ultimately manifest 
themselves in such a way.)

\noindent
ii) He may assume that looking at the conditions of the experiment only at the level of 
averages is insufficient. He must start to search which of the exterior measurements 
monitoring the conditions of his experiment he will have to analyse at the level of clicks. Then 
he will have to perform the experiment anew, where he now bins the data according to 
which of the possible outcomes are occurring at that other measurement. Thus he must 
observe correlations of clicks at two sites. If this is still not sufficient to obtain the full range 
from 0 to 1 for the various {\it coincidence probabilities} by adjusting the experimental 
conditions, he must postulate the need to look for yet another one of the measurements of the 
conditions and look at it at the level of clicks. He would then have triple-coincidence 
probabilities. As a general rule he must integrate so many of the condition-monitoring 
measurements and take them into account at the level of clicks, until the observed multi-
coincidence probabilities can be tuned to any value between 0 and 1 by changing proper 
parameters (in turn monitored by exterior measurements, but not at the level of clicks). Thus 
he is lead to the overall view.

We will now analyse the experiment of Fig.10 by adopting the overall view and assuming that 
no further exterior conditions need to be monitored at the level of clicks. Suppose we perform 
$N$ trials and we obtain the numbers of concident firings $L_{11}$, $L_{12}$, 
$L_{21}$, and $L_{22}$.  The first and the second subscript refer to the detector which 
fired at site $A$, and at site $B$, respectively. These numbers obey the constraint
\be
L_{11}+L_{12}+L_{21}+L_{22} = N.
\ee
With this we can estimate the unknown probabilities of the four possible outcomes. Their 
best estimates and the respective uncertainties are (see eq.(3) and eq.(6))
\be
p_{jk} = \frac{L_{jk}}{N},
\ee
\be
\Delta p_{jk} = \sqrt{\frac{p_{jk}(1-p_{jk})}{N}},
\ee
where $j$ and $k$ can each be  1 or 2.
And using eq.(24) we also obtain the associated {\it physical quantities} $\beta_{jk}$ 
as
\be
\beta_{jk} =\left(\sqrt{1-\frac{L_{jk}}{N}} + i 
\frac{L_{jk}}{N}\right) e^{-i\varphi_{jk}},
\ee
whose uncertainties are invariant under different results, as required (eq.(25)):
\be
\Delta \beta_{jk} = \frac{1}{2\sqrt{N}}.
\ee
Clearly, we have
\be
\left|\beta_{11}\right|^2 + 
\left|\beta_{12}\right|^2 + 
\left|\beta_{21}\right|^2 + 
\left|\beta_{22}\right|^2  = 1,
\ee
so that the experiment yields three independent random variables. From the previous section 
we know that the phases $\varphi_{jk}$ are arbitrary as long as we are not basing 
predictions on them. Since a common phase can be extracted, we have three free phases 
whose values cannot be determined in the experiment, but which we must consider as 
measurable, in principle. Altogether we must thus postulate six independent possibilities of 
influencing the outcomes in an experiment of the type of Fig.10. We note that quantum theory 
comes to the same conclusion: It also requires three independent complex probability 
amplitudes, hence six real numbers, to describe the situation. The result of the observation 
can now be written as a complex vector of the physical quantities $\beta_{jk}$. Setting 
$\varphi_{22}=0$ ist looks like this:
\be
\left( \begin{array}{c}
\beta_{11}\\
\beta_{12}\\
\beta_{21}\\
\beta_{22}\\
\end{array}
\right) =
\left(\begin{array}{l}
\left(\sqrt{1-\frac{L_{11}}{N}} + i \frac{L_{11}}{N}\right) e^{-
i\varphi_{11}}\\
\left(\sqrt{1-\frac{L_{12}}{N}} + i \frac{L_{12}}{N}\right) e^{-
i\varphi_{12}}\\
\left(\sqrt{1-\frac{L_{21}}{N}} + i \frac{L_{21}}{N}\right) e^{-
i\varphi_{21}}\\
\sqrt{\frac{(L_{11}+L_{12}+L_{21})}{N}} + i \frac{N-L_{11}-L_{12}-
L_{21}}{N}\\
\end{array}
\right).
\ee
The three free random variables are $L_{11}$, $L_{12}$, $L_{21}$ and the three free 
phases are $\varphi_{11}$, $\varphi_{12}$, $\varphi_{21}$.

The result vector indicates in no way that it represents coincidence outcomes at pairs of 
detectors at two different sites. It could equally well be the result vector of a probabilistic 
experiment where one trial can give four different outcomes at just {\it one} site. 
Therefore, whatever functional dependencies of the outcomes on the experimental conditions 
we observe, they are neither restricted nor expanded by the fact that the data are collected 
as coincidences of two clicks at two detectors which are distant from each other. The 
Einstein-Podolsky-Rosen paradoxon rests on the acceptance of "distance" and "spatiality" as 
we know them from everyday life as unquestioned primitive notions \cite{Zeilinger}. 
Already special and general relativity have shown the need to scrutinize these notions, 
although they have made no attempt to found them on more elementary ideas. Indeed, an 
understanding of "distance", "orientation", "space" (of whatever dimension) from pre-
geometric concepts seems a formidable task. But it is of utter importance if we take the 
common or the ultimate observer of Fig.1 seriously as the bare minimum needed to start doing 
empirical science. 

One can now analyse the experiment of Fig.10 when measuring the coincidence probabilities 
as a function of various parameters. But we will not really find anything different from the 
quantum theoretical description of the possible measurements on two correlated spin-1/2 
systems, although we would understand the emerging structures as consequences of the 
method how the observer gains information at the level of clicks. Therefore, we shall leave 
this as a possibly entertaining exercise to the reader.

\section{Discussion}

Several of the points presented in this paper deserve further comments. The first is our 
unquestioned acceptance that clicks just occur. What is a click, when looked at more 
closely? Philosophically, it is the transition from not-knowing to knowing which of several 
possibilities has become actual. For the human individual it is the advance forward in time by 
one small step, experiencing what this moment is filled with, and {\it naming} this 
experience. A name is, of course, the statement of exactly one out of finitely many 
possibilities. This connects it to technical measurement, which consists in recording which of 
finitely many pointer positions is found. Why in both cases only a finite amount of information 
is captured in finite time is not clear. Perhaps, because mind balks when trying to think the 
opposite (at least my mind does).

A second point is whether our search for invariants in probabilistic experiments must be 
restricted to probabilistic experiments that are in some sense "fundamental". What is the 
difference of observing the fraction of red cars in a city, to pick up on our initial doubts, to 
observing the fraction of counts in a particular detector? Well, the probabilistic measurement 
of the fraction of red cars in a city is only superficially a yes-no experiment. In fact it is a 
probabilistic experiment with a huge number of possible outcomes, and we just subsume one 
group of the outcomes into "red" and the other group into "not red". While this may have 
become clear in our analysis of correlations, let us look at it again in an example.

Data taken from a "fundamental" scattering experiment where a particle can end up in one of 
$M$ detectors look no different to the data obtained by letting balls bounce through an array 
of obstacles so that each ball can end up in one of $M$ different baskets. Both will conform 
to the multinomial distribution of order $M$. In both cases we can observe the probabilities 
as a function of various parameters. The difference is that in the ball experiment we are 
actually not distinguishing only $M$ different outcomes, but many, many more. A ball hitting 
a specific obstacle and going to basket 1 constitutes a different experiment to a ball not 
hitting this obstacle but still ending up in basket 1. We have been sloppy and not monitored 
the conditions properly! So we can start to do all kinds of measurements on the obstacles 
while the balls are bouncing through, and select out only those cases, where we are 
convinced that they were done under the same conditions of the obstacles, possibly the same 
temporal evolution of the conditions of the obstacles while a ball is bouncing through. But 
then we find that one ball in basket 1 may have come to lie there differently from the next ball 
in basket 1. Again we have been sloppy! We must now be even more discriminate and select 
only those cases, where the ball ends up in exactly the same position and same orientation in 
a basket. This is still not enough. We can still make many more distinctions in a trial than just 
the $M$ possible outcomes, e.g. we could screen the balls for the most minute differences in 
weight, we could record the patterns of light that are reflected from a ball while bouncing 
through, etc. But all these externally testable conditions must be the same in each trial, 
before we can say that we are doing a fundamental probabilistic experiment. If we manage to 
do that with balls, then we are justified in forming predictions on the basis of our analysis. 
Our conclusion is here no different to that of quantum theory \cite{Zeilinger2}.

This raises the question, whether fundamental probabilistic experiments are possible at all. 
The strict answer is {\it no},  because an observation can never be repeated. As we have 
already noted, an experimenter cannot monitor the whole world and ensure all conditions are 
the same in the successive trials of a probabilistic experiment. Epistemologically, the 
minimum difference is the record of the previous trials. An experimenter can just try to find 
out {\it which} conditions must be controlled and {\it which} can be neglected. In fact, 
this is how we learn about the world, and it gives us a basis for categorizing it! It is very likely 
that further analysis along the principle guideline of this paper that, through observation our 
knowledge about the world can only increase, general rules can be found of how to separate 
relevant from irrelevant conditions. Presumably this  is related to the concept of "distance", 
whose origin clearly lies in human muscular self-experience, and which must therefore be 
cracked. But as we have not addressed how probabilistic observation is connected to a 
number of terms which are similarly up for disection, but are indispensible for {\it doing} 
physics, terms like "system", "space", "interaction", etc., we cannot say anything more 
concrete.

Let us now comment on the consequences of the fact that the physical quantitity $\beta$, 
which we found associated with a measurable probability, is a complex variable.
We have noted that the prediction function $\beta'(t)$ is only uniquely determined when 
we assign specific values to the phases of the observed physical quantities $\beta_j$ 
(eq.(35) and Fig.9). More correctly, we must assign specific phase {\it differences} 
$\varphi_j-\varphi_k$ between any two observed $\beta_j$, $\beta_k$. They 
influence our predictions, which are testable, so that we must take these phase differences 
seriously as measurable quantities. This implies that, with each observable probability, which 
is just {\it one} number, we must associate {\it two measurable} numbers! One of 
these numbers is obtained by the observed relative frequency $L/N$, but the other is buried in 
the conditions and only its relative value is testable.  Nevertheless, as a general statement we 
find that, {\it a measurable probability is governed by two conditions, which can be 
controlled separately}. Of course, knowing quantum theory, this comes as no surprise. But 
our analysis started out by avoiding pre-conceived notions about the world, so that this finding 
has another significance. We can dare to speak of how a probability is related to the world. If 
we call an observable probability an {\it elementary system}, as has recently been 
advocated \cite{Caslav}, this turns into the visual imagery that an elementary system has 
a relation to the conditions of the world, which is given by two parameters. A closer 
inspection reveals, and actually Fig.5b is already suggestive of it, that these two parameters 
are equivalent to the polar coordinates fixing the direction of a unit vector in 3-dimensional 
cartesian space! Thus we can say that an observable probability, or an elementary system, 
has a controllable {\it orientation} with respect to the world, {\it as if} this world were 
3-dimensional. Of course, "distance" or "location" do not yet exist here. A similar connection 
between the 3-dimensionality of space and the fact that quantum theory assigns a complex 
two-component spinor to any yes-no problem has been emphasized by Weizs\"acker 
\cite{Weizs.}, and also expounded by Lyre \cite{Lyre2}. An interesting proof was 
sketched by Penrose \cite{Penrose} and worked out by Moussouris 
\cite{Moussouris}. The latter assigns an angle between any two multilevel quantum 
systems from the inner product of the angular momentum operators and then shows that the 
simultaneous angles of many such systems are compatible with presenting each system as a 
vector in 3-dimensional space. There, too, the concepts of "distance" and "location" are 
secondary.

Our method of constructing predictions requires a further remark. We have applied it only to 
predict probabilities when the experimental control parameter $t$ was set {\it between} 
values at which measurements have been performed. How shall we form predictions in cases, 
where we want to say something about the outcome probabilities to be expected in a {\it 
different} experimental setup than the one from which we obtained data for the prediction? 
What laws should we apply then? As a general rule, it must also be {\it linear laws}, just 
like those that we have used, because this ensures that more information obtained in the 
original experiment will lead to higher accuracy of the prediction for the other experiment. 
Take the double slit experiment as an example. We measure the probability of a particle to hit 
a certain small region on a screen behind the double slit. First we do a measurement with the 
left slit closed, then one with the right slit closed. Now we want to predict the probability 
when both slits are open. Here, we have no easy parameter $t$ with which to label the three 
experiments, so that the prediction would again be a kind of interpolation. Nevertheless, the 
{\it belief} that the two one-slit experiments have a lawful relation to the case when both 
slits are open leads us to apply a linear combination of the two respective quantities 
$\beta$. As one can verify, this gives the correct result, except for an unknown phase.

Finally, let us address the testability of our method of inferring physical quantities and forming 
predictions. We have already noted that we should expect any measurable probability to be 
adjustable between 0 and 1 --- no matter how many coincident clicks constitute an event. The 
practical difficulty may lie in finding the experimental conditions which one must tune to 
achieve this. Also, the method always refers to not yet controlled conditions if a test of a 
predicted probability disproves the prediction. It does not do so in "physical" terms, it only 
insists that there must be something in the conditions which has so far eluded our control 
while doing the trials. And formally it immediately offers a way to incorporate this 
uncontrolled condition: Its says that the trials so far must have happened under $R$ different 
conditions, where $R$ is an initially unknown integer. If we only manage to do the trials 
under any of these $R$ conditions, but always the same one, then our observed probability 
will become tunable between 0 and 1.  Effectively, it tells us to go to the next higher level of 
coincidence! This can lead to a regress without end. Thus there is no criterion of how to 
falsify this method. This is a point of worry, because our method seems to be identical to the 
quantum theoretical Hilbert space formalism. Quantum theory also just adds a few extra 
dimensions to Hilbert space and says they must have been floating during the experiment, 
otherwise we would have observed what it predicts. Consequently, quantum theory might be 
the frame from which there is no escape, and here I mean the Hilbert space formalism and not 
so much the physico-muscular lyrics around it. Of course, this conclusion is only warranted as 
long as we believe in the two tenets of empirical investigation on which we based our 
analysis:

\noindent
i) Any observation has finitely many possible outcomes and one just happens.

\noindent
ii) Only the probability of an outcome is meaningful.

\section {Conclusion}

In this paper we characterized the physicist's situation as being engulfed by a stream of 
observed data, where some data signify the conditions and others the outcomes of 
experiments. The most notable feature of this situation is that information increases with 
each observed bit, because with it the experimenter knows the answer to a particular yes-no 
question, while he had at best a guess before the observation. (And contrary to a widely held 
misconception, there are {\it no} sure guesses. New information {\it always} counts.) 

We argued that probability is the proper tool for extracting meaning, because it expects the 
least information from the data. Then we heeded Born's advice to simply look for invariants. 
This lead us eventually to complex probability amplitudes as random variables with the 
highest degree of invariance: Their uncertainty strictly decreases with each observation and 
is independent of the experimental result. A sequence of probabilities observed as a function 
of a tunable condition results in other quantities to represent probabilistic information, which 
have the additional symmetry of being independent of that condition. A complex probability 
amplitude is then seen as a linear superposition of these quantities.

Predictions for observable probabilities must similarly be formed by linear addition of complex 
probability amplitudes from other observed probabilities, in order to ensure that the prediction 
becomes more accurate with more observations as input, but is independent of the results of 
these experiments. In case a prediction fails, rather than doubting the law for making 
predictions, the search for invariants lead us to postulate that the experiments on whose data 
the predictions were based, must have been ill controlled, and that this can be remedied by 
going to a higher level of observing coincidence probabilities. As a consequence we obtain a 
description of the observer's world, {\it as if it were the Hilbert space of quantum theory}.

An insight of our analysis is that a probability, and more correctly a probability amplitude, is 
related to experimental conditions in two independent ways, so that one can see it {\it 
oriented with regard to the experimental conditions} as a unit vector in 3-dimensional 
cartesian space. This is, of course, ontological language. But the finding indicates a 
connection between the assumption of intrinsic probability and the unexplained effectiveness 
of believing to exist in a 3-dimensional world.

\section{Acknowledgment}
I am thankful for discussions on topics sometimes closer and sometimes more remote to the 
one presented here, but in my view always aimed at increasing our predictive power, 
discussions which I enjoyed with C. Brukner, Z. Hradil, G. Krenn and K. Svozil.

\pagebreak

\end{document}